%% file: 25_MFT_QuantumChains.tex
\documentclass{iopjournal}

\usepackage{amsmath}
\usepackage{float}

\usepackage[colorlinks=true, allcolors=blue]{hyperref}

\def\newblock{\hskip .11em plus .33em minus .07em}

\def\op#1{\textsf{\textbf{#1}} }
\def\nbOne{\hbox{{\bf 1$\!\!$l}}}
\def\eqref#1{(\ref{#1})}
\def\be{\begin{equation}}
\def\ee{\end{equation}}
\def\bea{\begin{eqnarray}}
\def\eea{\end{eqnarray}}
\def\nnb{\nonumber}
\def\bpm{\begin{pmatrix}}
\def\epm{\end{pmatrix}}

\begin{document}

\articletype{Paper} 

\title{Mean-field theory for quantum spin chains}

\author{Myl\`ene Martirosyan$^1$, Astrid Monin-Baroille$^1$, Le\"\i la Moueddene$^{2,3}$\orcid{0009-0001-7588-3835}, Mohammed M. Shabat$^{4,5}$\orcid{0000-0002-5382-9061} and Bertrand Berche$^{1,3,*}$\orcid{0000-0002-4254-807X}}

\affil{$^1$Laboratoire de Physique et Chimie Th\'eoriques, Universit\'e de Lorraine -- CNRS, Nancy, France}

\affil{$^2$Applied Theoretical Physics – Computational Physics
Physikalisches Institut, Albert-Ludwigs-Universit\"at Freiburg
Fribourg-en-Brisgau, Germany}

\affil{$^3$ $L^4$ collaboration Leipzig, Lorraine, Lviv, Coventry}

\affil{$^4$Islamic University of Gaza, IUG, Gaza, P.O.Box 108
Gaza Strip, Palestinian Authority}

\affil{$^5$Mathematics, Physics, and Electrical engineering Department,
Northumbria University, Newcastle upon Tyne NE1 8ST,  The U.K}

\affil{$^*$Author to whom any correspondence should be addressed.}

\email{bertrand.berche@univ-lorraine.fr}


\begin{abstract}
Mean-field theory is a widely used approximation for describing phase transitions, particularly effective above the upper critical dimension. Its origins can be traced back to the van der Waals theory of the liquid-gas transition and Weiss’s molecular field theory of the paramagnetic–ferromagnetic transition. However, it was Lev D. Landau who provided a unifying and general framework applicable to a broad class of physical systems. The mean-field approach typically involves neglecting thermal fluctuations, which is a reasonable assumption in many classical contexts. However, its application to quantum phase transitions at zero temperature is less common. The aim of this short pedagogical paper is to explore and clarify the use of the mean-field approach in the less familiar domain of quantum phase transitions. We specifically consider the Ising model and the Blume-Capel model.
\end{abstract}

\section{Introduction}

A particularly elegant and fruitful approach to studying thermal fluctuations in classical systems with interacting degrees of freedom in $d+1$ dimensions consists in mapping the problem onto one involving quantum fluctuations in a corresponding system with quantum degrees of freedom in $d$ spatial dimensions. This powerful correspondence -- reviewed in detail by Kogut~\cite{RevModPhys.51.659}, see also Fradkin and Susskind~\cite{PhysRevD.17.2637}, the book of Le Bellac~\cite{lebellac1998des} or Berche and L\'opez~\cite{Berche_2006} -- can be understood as a reinterpretation of the transfer matrix formalism for classical statistical systems, especially in the so-called Hamiltonian limit, where the transfer matrix acquires a simplified, quasi-Hamiltonian structure.

We will exploit this formal analogy between classical statistical mechanics and quantum mechanics~\cite{Berche2025} to reformulate the classical 2d Ising model as a 1d quantum system -- specifically, the Ising chain in a transverse field and the Blume-Capel quantum chain.

 To this end, we ``slice'' in the so-called time direction the energy of a given configuration of classical spin variables $\{s_{n,t}=\pm1\}$, with $1 \le n \le N$, $1 \le t \le T$, into a sum over the discrete time direction:
\begin{equation}
-\beta E[\{s_{n,t}\}] = -\beta \sum_t H[\sigma_n, \sigma'_n],
\end{equation}
where $t$ plays the role of (discrete) time, $n$ denotes the (remaining) space direction, and $\{\sigma_n\}$ and $\{\sigma'_n\}$ represent the spin configurations at times $t$ and $t+1$, respectively. This means that  $\{\sigma_n\}$ encodes the  $\{s_{n,t}\}'s$ and $\{\sigma'_n\}$ encodes the  $\{s_{n,t+1}\}'s$.  The Boltzmann weight associated with each time slice is then identified with a matrix element of a transfer operator $\textsf{\textbf{T}}$ acting on a quantum Hilbert space:
\begin{equation}
e^{-\beta H[\sigma_n, \sigma'_n]} = \langle \sigma' |\textsf{\textbf{T}} | \sigma \rangle.
\label{Eq_BoltzmannIMTM}
\end{equation}

The question we address here is how to build the transfer operator $\textsf{\textbf{T}}$ of the 1d quantum chain from the knowledge of the energy of the classical configurations $E[\{s_{n,t}\}] $. We will illustrate this procedure in the case of the Ising model~\cite{Ising1925,doi:10.1142/14565}, before dealing with the spin 1 Blume-Capel model~\cite{PhysRev.141.517,CAPEL1966966}.

These models are of course very well known and the first one in particular has been intensively studied, but we are not aware of any mean-field treatment, at least in pedagogical journals, for such quantum systems if we except for the work of Os\'acar and Pacheco~\cite{Osacar_2017} where the properties of the ground state of the Ising chain in a transverse magnetic field
is studied via a
Bethe–Peierls method. Therefore, we believe important to show how mean-field theory provides very interesting approximate results for otherwise extremely delicate quantum many-body systems.

\section{From classical to quantum spin models}
\subsection{Ising model}

Let us now pursue this program. We begin with the energy of a spin configuration on a square $N\times T$ lattice,
\begin{eqnarray}
-\beta E[s_{n,t}] &=& K_s\sum_{t=1}^T\sum_{n=1}^N s_{n,t}s_{n+1,t}
+ K_t\sum_{t=1}^T\sum_{n=1}^N s_{n,t}s_{n,t+1}\nonumber\\
  &=& \sum_t \Bigl[\sum_n \Bigl(
  K_s s_{n,t}s_{n+1,t} - \frac{1}{2}K_t (s_{n,t}-s_{n,t+1})^2
  \Bigr)\Bigr] + \hbox{const}.\label{Eq_EnergyTot}
  \end{eqnarray}
The degrees of freedom are the  classical spin variables $s_{n,t}=\pm 1$.
 They interact only with the nearest neighbouring spins via anisotropic ferromagnetic couplings $\beta J_i=K_i$, $i=s,t$ along the two directions of the lattice. The minus signs make the system ferromagnetic for $J_i>0$.
 This model is known as the two-dimensional Ising model for the square lattice.
  
  The relation \eqref{Eq_EnergyTot} naturally identifies inside the square brackets the ``sliced'' Hamiltonian
  \begin{equation}
  -\beta H[\sigma_n,\sigma'_n]
  =\sum_n \Bigl(
  K_s s_{n,t}s_{n+1,t} - \frac{1}{2}K_t (s_{n,t}-s_{n,t+1})^2\Bigr).
 \label{Eq_HamClass}
  \end{equation}

To write Eq.~\eqref{Eq_BoltzmannIMTM}, we must introduce the appropriate quantum states:
\begin{eqnarray}
&&|\sigma\rangle \equiv |\sigma_1,\sigma_2,\dots,\sigma_N\rangle ,\quad \sigma_n=s_{n,t}=\pm 1,\ \forall n,\\
&&|\sigma'\rangle \equiv |\sigma'_1,\sigma'_2,\dots,\sigma'_N\rangle ,\quad \sigma'_n=s_{n,t+1}=\pm 1,\ \forall n.
\end{eqnarray}
Acting on these states, we define the multi-site Pauli operators (here in the basis in which $\pmb\sigma_x$ is diagonal and $\pmb\sigma_z$ is a flipping operator)
\begin{eqnarray}
{\pmb \sigma}_x(n)&=&\nbOne_1\otimes\cdots\otimes\nbOne_{n-1}\otimes
\begin{pmatrix}1&0\\ 0&-1\end{pmatrix}_n
\otimes\nbOne_{n+1}\otimes\cdots\otimes\nbOne_N,\\
{\pmb \sigma}_z(n)&=&\nbOne_1\otimes\cdots\otimes\nbOne_{n-1}\otimes
\begin{pmatrix}0&1\\ 1&0\end{pmatrix}_n
\otimes\nbOne_{n+1}\otimes\cdots\otimes\nbOne_N,
\end{eqnarray}
which act as
\bea
{\pmb \sigma}_x(n)|\sigma_1,\dots,\sigma_n,\dots,\sigma_N\rangle  &=& \sigma_n |\sigma_1,\dots,\sigma_n,\dots,\sigma_N\rangle ,\\
{\pmb \sigma}_z(n)|\sigma_1,\dots,\sigma_n,\dots,\sigma_N\rangle  &=& |\sigma_1,\dots,-\sigma_n,\dots,\sigma_N\rangle ,
\eea
leaving unchanged all spin variables $\sigma_{m\not=n}$.

Since the transfer matrix cannot be read off directly, we classify row configurations ${\sigma}$ at time $t$ and ${\sigma'}$ at time $t+1$ according to the number of spin flips between $|\sigma\rangle$ and $|\sigma'\rangle$:
\bea
&|\sigma'\rangle \equiv |\sigma_{0\,\rm flip}\rangle \quad&\hbox{all spins unchanged, i.e.}\ \sigma'_n=\sigma_n,\ \forall n,\\
&|\sigma'\rangle \equiv |\sigma_{1\,\rm flip}\rangle \quad& \hbox{exactly one spin flipped, i.e.}\ \exists !\, n \ |\ \sigma'_n=-\sigma_n\\
&|\sigma'\rangle \equiv |\sigma_{2\,\rm flips}\rangle \quad &\hbox{exactly two spins flipped, i.e.}\ \exists !\, (n,m) \ |\ \sigma'_n=-\sigma_n\ \hbox{and}\ \sigma'_m=-\sigma_m\\
&\dots\nonumber
\eea
and we associate to each case an operator $\op T_{k\,\rm flips}$ generating the corresponding Boltzmann weight. Care must be taken to ensure that the operator associated with $k$ flips does not contribute to sectors with a different number of flips.

For instance, for zero flips, all terms $s_{n,t}-s_{n,t+1}=0$ in \eqref{Eq_HamClass} and one has
\begin{equation}
e^{-\beta H[\sigma_n,\sigma_{0\,\rm flip}]}
=e^{\sum_n K_s s_{n,t}s_{n+1,t}}
=\langle \sigma_{0\,\rm flip}|\op T_{0\,\rm flip}|\sigma\rangle ,\label{Eq_BoltzmannIMTM0flip}
\end{equation}
leading to the identification of the operator $\op T_{0\,\rm flip}$ which does the job
\begin{equation}
\op T_{0\,\rm flip}=
\exp\Bigl(K_s\sum_n{\pmb\sigma}_x(n){\pmb\sigma}_x(n+1)\Bigr).
\end{equation}

For one flip, we require
\begin{eqnarray}
e^{-\beta H[\sigma_n,\sigma_{1\,\rm flip}]}
&=& e^{\sum_n\bigl(K_s s_{n,t}s_{n+1,t}-\frac{1}{2}K_t(s_{n,t}-s_{n,t+1})^2\bigr)}\nonumber\\
&=& e^{K_s\sum_n s_{n,t}s_{n+1,t}} e^{-2K_t}\nonumber\\
&=&\langle \sigma_{1\,\rm flip}|\op T_{1\,\rm flip}|\sigma\rangle ,\label{Eq_BoltzmannIMTM1flip}
\end{eqnarray}
which allows to identify
\begin{equation}
\op T_{1\,\rm flip}=
\exp\Bigl(K_s\sum_n{\pmb\sigma}_x(n){\pmb\sigma}_x(n+1)\Bigr)
e^{-2K_t}\sum_n{\pmb\sigma}_z(n).
\end{equation}
Checking a simple example confirms this result: the $\sum_n{\pmb\sigma}_z(n)$ term selects exactly the flipped spin, for which it contributes a factor $e^{-2K_t}$ while unflipped spins do not contribute.  The spatial interaction term has vanishing matrix elements between one-flip states and can be omitted in the strict quantum limit. We nevertheless keep this term to allow a later factorisation.

The two-flip contribution follows directly,
\begin{equation}
\op T_{2\,\rm flips}=
\exp\Bigl(K_s\sum_n{\pmb\sigma}_x(n){\pmb\sigma}_x(n+1)\Bigr)
e^{-4K_t}\sum_{n\neq m}{\pmb\sigma}_z(n){\pmb\sigma}_z(m),
\end{equation}
and in general the $k$-flip term carries a factor $e^{-2kK_t}$ and a product of $k$ distinct flip operators. Collecting all terms gives the full transfer matrix:
\begin{equation}
\op T=
\exp\Bigl(K_s\sum_n{\pmb\sigma}_x(n){\pmb\sigma}_x(n+1)\Bigr)
\Bigl(\nbOne_{2^N}
+ e^{-2K_t}\sum_n{\pmb\sigma}_z(n)
+ e^{-4K_t}\sum_{n\neq m}{\pmb\sigma}_z(n){\pmb\sigma}_z(m)+\dots\Bigr).\label{Eq_TMexpanded}
  \end{equation}
  This may be written in the standard form $\op T=e^{-\epsilon \op H}$, thereby defining an effective quantum Hamiltonian $\op H$.

Although extracting $\op H$ is still nontrivial, the expression simplifies in the extreme anisotropic limit $K_s\to 0$, $K_t\to\infty$, with the parametrization
\begin{equation}
K_s=\lambda e^{-2K_t}, \qquad \lambda=O(1).\label{Eq_ExtrAnisLimit}
\end{equation}
Expanding the spatial term in powers of $K_s$ yields
\begin{eqnarray}
\exp\Bigl(K_s\sum_n{\pmb\sigma}_x(n){\pmb\sigma}_x(n+1)\Bigr)
&=& \nbOne_{2^N} + K_s\sum_n{\pmb\sigma}_x(n){\pmb\sigma}_x(n+1) \nonumber\\
&&+ \frac{1}{2}K_s^2\sum_{n,m}{\pmb\sigma}_x(n){\pmb\sigma}_x(n+1){\pmb\sigma}_x(m){\pmb\sigma}_x(m+1) + O(K_s^3),
\end{eqnarray}
so that the transfer matrix becomes
\bea
\op T&=&
\Bigl(\nbOne_{2^N}+K_s\sum_n{\pmb\sigma}_x(n){\pmb\sigma}_x(n+1)+\frac{1}{2}K_s^2\sum_{n,m}{\pmb\sigma}_x(n){\pmb\sigma}_x(n+1){\pmb\sigma}_x(m){\pmb\sigma}_x(m+1)+\dots\Bigr)\nonumber\\
&&\times\Bigl(e^{-2K_t}\sum_n{\pmb\sigma}_z(n)
+ e^{-4K_t}\sum_{n\neq m}{\pmb\sigma}_z(n){\pmb\sigma}_z(m)+\dots\Bigr).
  \eea
  Inserting \eqref{Eq_ExtrAnisLimit} and taking $\epsilon=e^{-2K_t}\to 0$, we find to lowest order
  \bea
  \op T &\simeq& \nbOne_{2^N} + e^{-2K_t}\Bigl(\lambda\sum_n{\pmb\sigma}_x(n){\pmb\sigma}_x(n+1)+\sum_n{\pmb\sigma}_z(n)\Bigr)+\dots \nonumber\\
  &=& \nbOne_{2^N}-\epsilon\op H,
  \eea
  from which the effective Hamiltonian emerges:
  \begin{equation}
  \op H_{\rm IM}=-\lambda\sum_n{\pmb\sigma}_x(n){\pmb\sigma}_x(n+1)-\sum_n{\pmb\sigma}_z(n).
  \label{Eq-H_IM}
  \end{equation}
  This is precisely the Hamiltonian of the quantum Ising chain in a transverse field.  The denomination of transverse field refers to the ${\pmb\sigma}_z(n)$ terms. Indeed, an ordinary magnetic field $h$ would couple to the local order parameter 
  ${\pmb\sigma}_x(n)$ and would contribute as $-h\sum_n{\pmb\sigma}_x(n)$ instead. Note that $ \op H_{\rm IM}$ is represented in the spin basis by a $2^N\times 2^N$ matrix for a chain of length $N$. When $h=0$, there exists an exact diagonalisation in free fermions~\cite{RevModPhys.51.659,karevski}.

\subsection{Blume-Capel model}

The Blume-Capel model is an extension of the Ising model to spin 1, $s_{n,t}=-1,0,1$ and a crystal field $\Delta$ which couples to the $s_{n,t}^2$.
 The   ``sliced'' Hamiltonian reads as
  \begin{equation}
  -\beta H[\sigma_n,\sigma'_n]
  =\sum_n \Bigl(
  K_s s_{n,t}s_{n+1,t} - \frac{1}{2}K_t (s_{n,t}-s_{n,t+1})^2
  -\Delta s_{n,t}^2
  \Bigr).
 \label{Eq_HamClassBC}
  \end{equation}
  The role of the crystal field is that of a chemical potential (the sign for $\Delta$ is conventional). Indeed, it controls the population of the states $s_{n,t}=0$ which are not coupled in energy through the ordinary $K_s-K_t$ couplings. When $\Delta=0$, this term disappears, leaving essentially only an Ising model with $s=\pm 1$ -- the $s_{n,t}=0$ states only contribute to the entropy there and are completely suppressed only in the limit $\Delta\to-\infty$ --, while $\Delta >0$ renders the $ 0$ state more favourable, contributing to the entropy of the system (snapshots of typical Monte Carlo configurations showing the role of the proliferation of the zero state can be found in Ref.~\cite{Moueddene_2025}.).

In the extreme anisotropic limit, the quantum Blume-Capel model becomes
\be
 \op H_{\rm BC}=-\lambda\sum_n{\op s}_x(n){\op s}_x(n+1)+D\sum_n{\op s}_x(n)^2-\sum_n{\op s}_z(n).
  \label{Eq-H_BC}
\ee
with
\begin{eqnarray}
{\op s}_x(n)&=&\nbOne_1\otimes\cdots\otimes\nbOne_{n-1}\otimes
\begin{pmatrix}1&0 & 0\\ 0 & 0 & 0\\ 0&  0&-1\end{pmatrix}_n
\otimes\nbOne_{n+1}\otimes\cdots\otimes\nbOne_N,\\
{\op s}_x(n)^2&=&\nbOne_1\otimes\cdots\otimes\nbOne_{n-1}\otimes
\begin{pmatrix}1&0 & 0\\ 0 & 0 & 0\\ 0&  0&1\end{pmatrix}_n
\otimes\nbOne_{n+1}\otimes\cdots\otimes\nbOne_N,\\
{\op s}_z(n)&=&\nbOne_1\otimes\cdots\otimes\nbOne_{n-1}\otimes
\frac{1}{\sqrt 2}\begin{pmatrix}0&1 & 0\\ 1 & 0 & 1\\ 0 & 1&0\end{pmatrix}_n
\otimes\nbOne_{n+1}\otimes\cdots\otimes\nbOne_N,
\end{eqnarray}
and $\Delta=D e^{-2K_t}\to 0$.
 In the following, we will restrict to the regime $D\ge 0$.

\section{Quantum mean-field for the Ising  chain}

We now turn to the mean-field approximation of the Hamiltonian~\eqref{Eq-H_IM}.  
The basic idea is to decouple spin--spin correlations by writing
\begin{equation}
{\pmb\sigma}_x(n) = \langle {\pmb\sigma}_x(n)\rangle + \delta{\pmb\sigma}_x(n),
\end{equation}
so that fluctuations are encoded in $\delta{\pmb\sigma}_x(n)$, with 
$\langle \delta{\pmb\sigma}_x(n)\rangle=0$.  

For the product of two neighboring spins we obtain
\begin{align}
{\pmb\sigma}_x(n){\pmb\sigma}_x(n+1)
&= \bigl(\langle{\pmb\sigma}_x(n)\rangle+\delta{\pmb\sigma}_x(n)\bigr)
   \bigl(\langle{\pmb\sigma}_x(n+1)\rangle+\delta{\pmb\sigma}_x(n+1)\bigr)\notag\\
&\simeq 
 {\pmb\sigma}_x(n)\langle{\pmb\sigma}_x(n+1)\rangle
+{\pmb\sigma}_x(n+1)\langle{\pmb\sigma}_x(n)\rangle
-\langle{\pmb\sigma}_x(n)\rangle\langle{\pmb\sigma}_x(n+1)\rangle,
\end{align}
where terms quadratic in the fluctuations 
$O(\delta{\pmb\sigma}_x(n)\delta{\pmb\sigma}_x(n+1))$ have been neglected.  

Assuming a spatially homogeneous system -- i.e. a chain with periodic boundary conditions (PBC) ${\pmb\sigma}_i(n+N)={\pmb\sigma}_i(n)$ --, we drop the site index and define
\begin{equation}
m=\langle{\pmb\sigma}_x\rangle,
\end{equation}
which plays the role of the order parameter.  
Substituting into the Hamiltonian gives the mean-field form
\begin{equation}
\op H_{\rm MF}=-\lambda\sum_n \bigl(2m\,{\pmb\sigma}_x-m^2\bigr)
-\sum_n{\pmb\sigma}_z.
\end{equation}
Thus, the mean-field Hamiltonian per site is
\begin{equation}
\frac{\op H_{\rm MF}}{N}=-2\lambda m\,{\pmb\sigma}_x-{\pmb\sigma}_z+\lambda m^2\nbOne=\op A +\lambda m^2\nbOne.\label{Eq_HMF}
\end{equation}

Solving this expression is an undergrad exercise.
The matrix form of the Hamiltonian depends on the representation chosen for the Pauli matrices, and here, in order to keep track of the $\sigma_x=\pm 1$ in the original Ising model, we go on working in   the basis in which ${\pmb\sigma}_x$ is diagonal, hence
\begin{equation}
N^{-1}(\op H_{\rm MF})_{\sigma_z}=-\begin{pmatrix}  2\lambda m & 1\\ 1& -2\lambda m \end{pmatrix}
+\lambda m^2\begin{pmatrix} 1 & 0\\ 0 & 1\end{pmatrix}.
\end{equation}
The first two terms in \eqref{Eq_HMF} describe a single spin in an effective magnetic field, $-{\vec B}_{\rm eff}\cdot\vec{\pmb\sigma}$ with
${\vec B}_{\rm eff}=(1,0,2\lambda m)$, leading automatically to the eigenvalues of  \eqref{Eq_HMF} 
\begin{equation}
E_\pm(m)=\lambda m^2\pm|\vec B_{\rm eff}|=\lambda m^2\pm \delta(m),\quad \delta(m)=\sqrt{1+(2\lambda m)^2}.\label{Eq_E+-}
\end{equation} 
The condition $\delta(m)^2-(2\lambda m)^2=1$ is useful.

The ground state energy is $E_{\rm GS}=E_-(m)$, and the corresponding normalised eigenstate can be parametrised according to
\begin{equation}
|\psi_{\rm GS}\rangle= \begin{pmatrix} \cos\theta\\ \sin\theta\end{pmatrix}.
\end{equation} 
Inserting this expression in the eigenvalue problem $N^{-1}\op H_{\rm MF} |\psi_{\rm GS}\rangle = E_{\rm GS}|\psi_{\rm GS}\rangle$ 
leads to 
\begin{equation}
\cos^2\theta=\frac{\delta(m)+2\lambda m}{\delta(m)},\quad 
\sin^2\theta=\frac{\delta(m)-2\lambda m}{\delta(m)}
\end{equation} 
and allows for the calculation of the average magnetisation $\langle {\pmb\sigma}_x\rangle_{\rm GS}$ per site:
\begin{eqnarray}
\langle {\pmb\sigma}_x\rangle_{\rm GS}&=&
(\cos\theta\ \sin\theta)\begin{pmatrix} 1 & 0\\ 0 & -1\end{pmatrix} \begin{pmatrix} \cos\theta\\ \sin\theta\end{pmatrix}\nonumber\\
&=&\cos^2\theta-\sin^2\theta.
\end{eqnarray} 
After simplifications, we obtain
\begin{equation}
\langle {\pmb\sigma}_x\rangle_{\rm GS}=\frac{2\lambda m}{\delta(m)}.
\end{equation}
Self-consistency demands that $\langle {\pmb\sigma}_x\rangle_{\rm GS}=m$ and, solving for $m$
we get
\be m^2(\lambda)=1-\frac{1}{4\lambda^2}.
\ee
It vanishes at $\lambda=\lambda_c=\frac 12$ and  for $\lambda\le \lambda_c$ -- the paramagnetic phase.

To investigate the universality class of the model, we analyse the behaviour in the vicinity of the phase transition (see Supplementary Material (SM) section 1, or Ref.~\cite{doi:10.1142/5376} and the whole series of books from this series).
Thanks to the scaling laws among critical exponents, the values for two of them are usually\footnote{The case of the first of these relations is a bit special in the sense that it contains the space dimensionality. It is called the hyperscaling relation and requires a special care, in particular in the mean-field regime~\cite{PhysRevLett.89.025703,doi:10.1142/9789819800827_0001,SciPostPhysLectNotes.60}. It is enough to know that $d$ must be fixed to the upper critical dimension there (see SM, sec. 1).} needed to recover all exponents
\bea 
&&\alpha=2-d\nu,\\
&&\alpha+2\beta +\gamma=2,\\
&&\beta(\delta-1)=\gamma,\\
&&(2-\eta)=\gamma/\nu.
\eea

When $\lambda\to\lambda_c=1/2$, there is a phase transition between a disordered phase at high values of $\lambda^{-1}$, where the ${\pmb\sigma}_z$ term dominates the Hamiltonian and produces spin flips, and a ferromagnetic ordered phase at small values of $\lambda^{-1}$, where ${\pmb\sigma}_x{\pmb\sigma}_x$ dominates, inducing order in $\sigma_x$\footnote{Note that the mean-field transition coupling $\lambda_c=1/2$ is not in agreement with the exact result $\lambda_c^{\rm exact}=1$, and in the same manner, the critical exponent of the order parameter does not agree with the exact $\beta_{\rm exact}=1/8$ ~\cite{RevModPhys.51.659,PFEUTY197079}.
}
\begin{eqnarray}
&m=0\quad&\lambda^{-1}>\lambda_c^{-1},\\
&m
\simeq 2\lambda_c(\lambda_c^{-1}-\lambda^{-1})^{1/2}\quad&\lambda^{-1}<\lambda_c^{-1},
\end{eqnarray}
i.e. a phase transition in the standard Ising mean-field universality class -- with a critical exponent $\beta_{\rm MFT}=1/2$ (see the SM, sec. 1 for the definition of the exponents) and a scaling field \be\tau=\lambda^{-1}-\lambda_c^{-1}\ee 
such that $m\simeq |\tau|^{\beta_{\rm MFT}}$ for $\tau<0$.
Since a second critical exponent is needed to fully have the universality class, we have to take the external magnetic field into account. The mean-field  Hamiltonian is now
\begin{equation}
\frac{\op H_{\rm MF}}{N}=-2\lambda m\,{\pmb\sigma}_x-{\pmb\sigma}_z-h{\pmb\sigma}_x+\lambda m^2\nbOne,\label{Eq_HMFh}
\end{equation}
the eigenvalues keep their form \eqref{Eq_E+-} with $\delta(m)$ now replaced by
\be
\delta(m,h)=\sqrt{1+(2\lambda m+h)^2}.
  \ee 
  The magnetisation becomes solution of
  \be 
  m =\frac{2\lambda m+h}{\sqrt{1+(2\lambda m+h)^2}}.
  \ee
At the critical value $\lambda_c$, $m_c=m(\lambda_c,h)$ obeys 
\be 
m_c^3\simeq 2h,
\ee
hence a critical exponent $\delta_{\rm MFT}=\frac 13$.

We are studying a $T=0$ (inverse temperature $\beta$ infinite), a so-called a ground-state quantum problem. As $\beta\to\infty$
\be
Z=\hbox{Tr}\, e^{-\beta \op H}=\sum_{k=\pm} e^{-\beta E_k}\to e^{-\beta E_-}
\ee
so that 
\be
-\frac 1\beta \ln Z = E_-
\ee
holds exactly.
The excited levels are exponentially suppressed and contribute nothing to the ground-state energy or to $m=\langle {\pmb\sigma}_x\rangle $, $\lambda$
 plays the role of a coupling driving a genuine quantum phase transition at $T=0$,
  not a thermal one.
  
Therefore, an expansion of the ground state $E_-$ leads to a variational Landau expansion of the free energy per site. It is equal to
\bea  
F&=&E_{\rm GS}=\lambda m^2-\sqrt{1+(2\lambda m+h)^2}\nonumber\\
&=& -1-2\lambda mh+\lambda (1-2\lambda)m^2+2\lambda^4 m^4 +O(m^6,h^2,\dots)\nnb\\
&=&f_0+\tfrac 12 r_0\tau m^4+\tfrac 14 u_0 m^4.
\eea  
This is indeed a standard $\phi^4-$Landau expansion (see SM, sec. 2) with positive constants $r_0$ and $u_0$ and a critical value $1-2\lambda_c=0$ where the coefficient of the quadratic term vanishes, in agreement with the previous analysis.

\section{Quantum mean-field for Blume-Capel chain}

The Blume--Capel model has experienced a revival of interest in recent years, mainly owing to its rich phase diagram. In the classical case, it features a second-order transition line between the ferromagnetic and paramagnetic phases, which terminates at a tricritical point and continues as a first-order transition line. Since its introduction, the model has been extensively studied in one~\cite{PhysRevE.76.021104,BHATTACHARJEE2026131968}, two~\cite{PhysRevE.73.036702,zierenberg_scaling_2017,Moueddene_2024,e26030221,PhysRevResearch.7.013214,jeon2026clusterdynamicsstayfastuntil}, and three dimensions~\cite{Hasenbusch_2010,PhysRevE.91.032126,PhysRevE.110.064144,MATARAGKAS2026131922}, as well as on the complete graph~\cite{10.1063/10.0036500}, using a variety of analytical and numerical approaches. More recently, quantum extensions have also been investigated, including the model in a transverse crystal field using variational approaches~\cite{CARVALHO2015240}, its quantum spin-$2$ variant using a variational approach~\cite{7gwr-n7vt}, the quantum $J_1$-$J_2$ Blume--Capel model using cluster mean-field theory~\cite{guerrero_quantum_2025}, and the quantum Blume--Capel model using mean-field theory and quantum Monte Carlo approaches~\cite{doi:10.1143/JPSJ.74.2957}.
The mean-field Hamiltonian per site is
\begin{equation}
\frac{\op H_{\rm MF}}{N}=-2\lambda m\,{\op s}_x+D{\op s}_x^2-{\op s}_z+\lambda m^2\nbOne
=\op B+\lambda m^2\nbOne
.\label{Eq_HMFb}
\end{equation}
\be
N^{-1}(\op H_{\rm MF})_{s_x}=\begin{pmatrix}
D-2\lambda m & -2^{-1/2} & 0 \\
-2^{-1/2} & 0 & -2^{-1/2} \\
0 & -2^{-1/2} & D+2\lambda m
\end{pmatrix}
+ \lambda m^2
\begin{pmatrix}
1 & 0 & 0\\
0 & 1 & 0\\
0 & 0 & 1
\end{pmatrix}.
\ee
Solwing this model requires the diagonalisation of $\op H_{\rm MF}$.

\subsection{Cardano solutions for the eigenvalue problem}

This is a real symmetric matrix; therefore we can use Cardano's real solutions. The determinant $P(b)=\hbox{\rm Det}\,(\op B-b\nbOne)$ reads as
\bea 
P(b)=-b^3+2Db^2-(D^2-1-4\lambda^2 m^2)b-D.\eea
Setting $b=z+2D/3$, we have to find the roots of 
$z^3-pz+q=0$ where 
$p(m) =1+4\lambda^2 m^2+D^2/3$, 
$q(m)=\frac D3(1-8\lambda^2 m^2)+2D^3/27$.

The three solutions of $P(b)=0$ are thus given (see SM, sec. 3), for $k=0,1,2$  by
\bea
b_k&=&z_k+2D/3,\\
z_k&=&\frac{2\sqrt{p}}{\sqrt 3}\cos\theta_k,\\
\theta_k&=&\frac{\varphi+2k\pi}{3},
\eea 
where $\varphi=\arccos (-X(m))$, $X(m)=(3\sqrt 3/2) (q/p^{3/2})$.
The eigenvalues of $N^{-1}\op H_{\rm MF}$ are eventually given by \be\epsilon_k(m)=b_k+\lambda m^2=z_k+2D/3+\lambda m^2.\ee

\subsection{Consistency check}

Let us first check the consistency of the results and look at the limit $D=0$. There, we simplify $p_0=1+(2\lambda m)^2$, $q_0=0$, $X_0=0$, $\varphi_0=\pi/2$. We deduce the three values of $\theta_k(D=0)=\pi/6$, $5\pi/6$ and $3\pi/2$ and identify the ground states with $k=1$, i.e. $y_{\rm GS}=-\sqrt{1+(2\lambda m)^2}$, 
\be \epsilon_{\rm GS}(m)=\lambda m^2 -\sqrt{1+(2\lambda m)^2}.\ee

We then obtain the self-consistency equation for $m$,
\be m=\frac{2\lambda m}{\sqrt{1+(2\lambda m)^2}}.\ee
It agrees with the previous results for the Ising model as it was announced when we discussed the limit $\Delta =0$ of the classical model.

 We can also note that in the limit $D\to \infty$, the quantity $X_\infty\to 1$, which implies $\varphi_\infty=0$ and therefore $\theta_0(D\to\infty)\to 0$. This proves that $\theta_0(D)$ varies between $\pi/6$ and $0$ for $0\le D\le \infty$, and this implies that the ground state is always given by $k=1$ for the whole interval of values of $D$ (see SM, sec. 3 for illustration).

\subsection{Weak crystal field limit}
The next question we may ask is whether the critical point shifts at higher, or at lower values of $\lambda$, when the crystal field $D\not=0$. We expect that non-zero value of $D$ softens the transition, since it hinders the implementation of the order, i.e. we expect that $\lambda_c^{-1}(D)\le  \lambda_c^{-1}(0)=2$.

To check this expectation, we analyse the magnetisation as given by 
\be 
\langle \op s_x\rangle_k=\frac{2A(m)B(m)}{A^2(m)+B^2(m)+(A^2(m)-B^2(m))^2},\label{Eq_mABetc}
\ee 
with $A(m)=D-b_k$ and $B(m)=2\lambda  m$
in the SM, sec. 3.
In the limit $m\to 0$ to linear order in the parameter $D$, we have
$p_D\simeq 1 $, $q_D\simeq \tfrac D3 $, from where we deduce $X_D\simeq \tfrac{\sqrt{3}}{2} D $, $z_1\simeq -1 -D/6$. It follows that
 $B=2\lambda m$ and $A\simeq 1+D/2$ and therefore 
 \be
 m=\frac{4A\lambda m}{A^2+A^4+4\lambda^2 m^2(1-2A^2)},
 \ee 
 leading to 
 \be 
 m^2=\frac{A(4\lambda - A - A^3)}{4\lambda^2(1-2A^2)}
 \ee 
 which vanishes at 
 \be
 \lambda_c(D)=\frac 12+\frac D2,
 \ee
 and confirms what we have announced concerning the direction of the shift.

\subsection{Free energy of the model}

Following the same strategy as in the Ising case, an expansion of the ground state $\epsilon_1$ leads to a variational Landau expansion of the free energy. It  is of the following form (see SM, sec. 3 for details)
\be
f(m)=\epsilon_1(m)=f_0+\tfrac 12 r m^2+\tfrac 14 u_4 m^4+\tfrac 16 u_6 m^6\label{Eq_LandauExp}\ee
with $u_6>0$. The  coefficient $u_4$ vanishes at the tricritical value of the crystal field 
\be D_{\rm tri}=\sqrt{\sqrt 5-2}\simeq 0.4859.\ee
For any $D\le D_{\rm tri}$, the coefficient $r$  vanishes at the critical value of the coupling 
\be
\lambda_c(D)=\frac{\sqrt{D^2+4}(\sqrt{D^2+4}+D)^2}{16}
\ee
leading to $\lambda_c^{-1}(0)=2$, $\lambda_c^{-1}(0.3)=1.467$, 
$\lambda_c^{-1}(D_{\rm tri})=1.201$ (details can be found in the SM, sec. 3).

\begin{figure}[!t]
 \centering
        \includegraphics[width=0.6\textwidth]{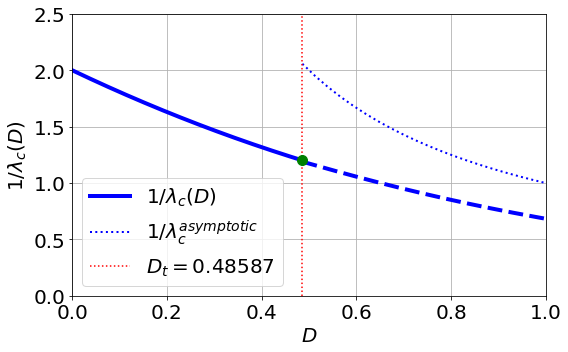}
 \caption{Phase diagram of the BC model. The vertical dotted line (red online) marks the tricritical value $D_{\rm tri}$. The  continuous line (blue online) at lower values of $D$ is the second order regime while the dashed line (blue online) at larger values is the first order transition. The dotted line above the tricritical point is the asymptotic $D\gg 1$ first-order transition coupling.}
\label{fig1}
\end{figure}

 The phase diagram in shown in Fig.~\ref{fig1}. The continuous line represents the second-order transition line, which is in the mean-field Ising model universality class, and ends at the value $D_{\rm tri}$ corresponding to the mean-field tricritical universality class. For higher values of the crystal field $D$, the transition becomes a first-order one, and we have determined the transition coupling in the vicinity of the tricritical point. It is shown in dashed line in Fig.~\ref{fig4} which also exhibits an asymptotic curve at large values of $D$ in dotted line. It is remarkable that the transition remains first-order until $D\to \infty$ in this case and that the line of first-order transition continues the second-order line and does not end abruptly at finite $D$ like in the finite-dimensional classical Blume-Capel model 
~\cite{Moueddene_2024,PhysRevE.110.064144,10.1063/10.0036500,moued2025}, even in the mean-field regime of the model at $d=3$ -- the upper critical dimension of the tricritical model.

\begin{figure}[t]
 \centering
        \includegraphics[width=1.\textwidth]{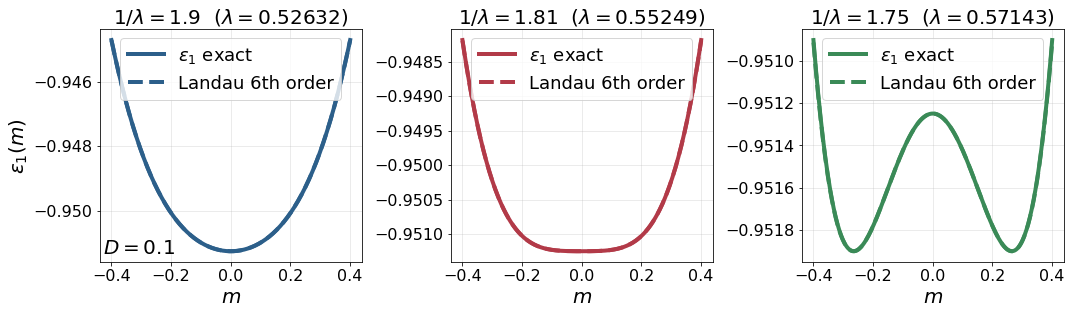}
         \includegraphics[width=1.\textwidth]{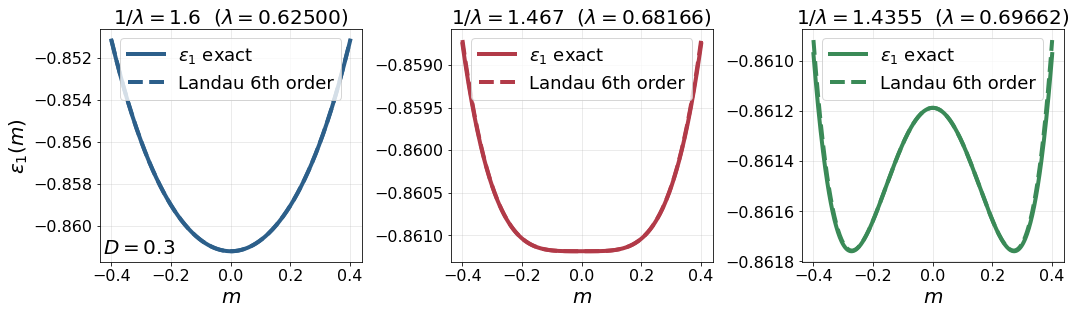} 
 \caption{Landau expansion of the free energy crossing the Ising universality class transition at $D=0.1$, 0.3. At the transition coupling, the free energy is flat.}
\label{fig2a}
\end{figure}

\begin{figure}[!ht]
 \centering
         \includegraphics[width=1.\textwidth]{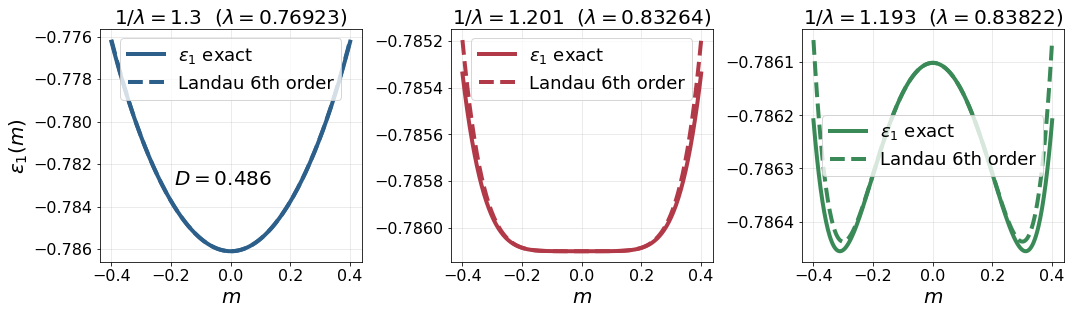} 
  \caption{Landau expansion of the free energy crossing the tricritical universality class transition at $D=0.486$. The plateau of the free energy at the transition is wider.}
\label{fig2b}
\end{figure}

\begin{figure}[ht]
 \centering
         \includegraphics[width=1.0\textwidth]{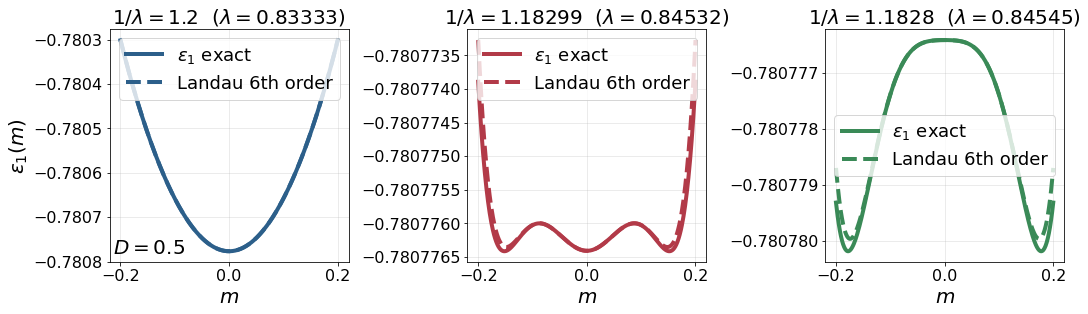}
         \includegraphics[width=1.0\textwidth]{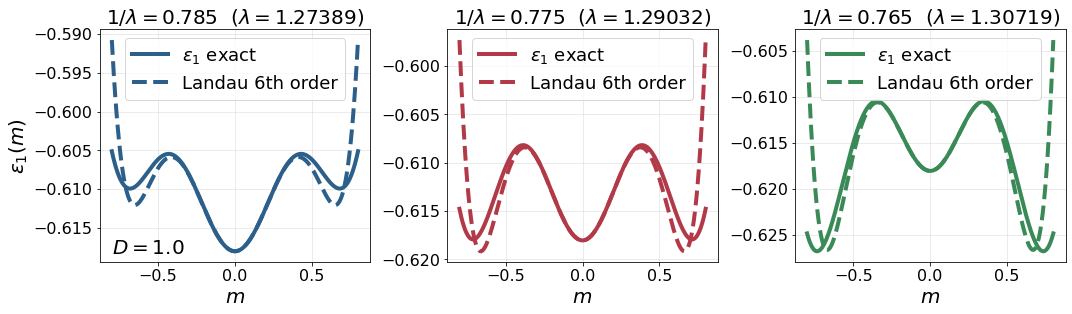}
 \caption{Landau expansion of the free energy crossing the first-order transition at $D=0.5$, 1.0. Note the difference of scales on the horizontal axis. At the transition, the discontinuity of the order parameter becomes more pronounced when we go deeper in the first-order regime (increasing $D$).}
\label{fig2c}
\end{figure}

The Landau expansions are shown, vs $m$, in Figs.~\ref{fig2a}, \ref{fig2b}, \ref{fig2c}, where we have used
\bea 
&&f_0=b_1(0),\\
&&r=2(\lambda + b_1'(0)),\\
&&u_4=2b_1''(0),\\
&&u_6=b_1'''(0),
\eea 
and
\bea 
&&b_1(0)=\frac{D-S}{2},\\
&&b_1'(0)=\frac{-16\lambda^2}{S(S+D)^2},\\
&&b_1''(0)=\frac{512\lambda^4\big[S(1-D^2)-D(D^2+3)\big]}{S^3(S+D)^5},\\
&&b_1'''(0)=\frac{-49152\,\lambda^6\big[D^7+D^6S+4D^5+2D^4S-8D^3-6D^2S-32D+4S\big]}{S^6(S+D)^8}.
\eea
with $S=\sqrt{D^2+4}$ (see SM, sec. 3 for the details of the calculation).

 In the second-order regime (Fig.~\ref{fig2a}), $D=0.1$ and 0.3, the free energy is essentially parabolic in the disordered phase (left) and has a double-minimum shape in the ordered phase (right). At criticality, the free energy flattens at the minimum. At the tricritical point  (Fig.~\ref{fig2b}) $D_{\rm tri}=0.486$, the structure is similar, but the flattening of the minimum at the transition is more pronounced. Eventually, in the first-order regime  (Fig.~\ref{fig2c}) $D=0.5$ and 1.0, at the transition there appears a three-minimum structure and the minimum at $m=0$ becomes strongly unstable in the ordered phase. 
 The deviation between the Landau expansion (dashed curves) and the exact expression (solid lines) of $\epsilon_1(m)$ is more pronounced in the first order regime, and is slightly visible at the tricritical point, but the agreement is excellent along the second-order line.

  \subsection{Strong crystal field limit}
 Above \(D_{\mathrm{tri}}\), the transition remains first-order up to $D\to\infty$. In the \(D\gg 1\) limit, the \(\op s_{z}\) flip term can be neglected, and the eigenvalues are given by $D-2\lambda m+\lambda m^2$, $\lambda m^2$, and $D+2\lambda m+\lambda m^2$. The two lowest-energy states are \(s_x=0\) and \(s_x=+1\). The \(s_x=0\) state favors $m=0$, where its energy is 0, whereas the \(s_x=1\) state favours $m=1$, where the energy equals $D-\lambda$. The two states exchange stability at the crossing point, which asymptotically corresponds to the transition \(\lambda_{\rm 1st}(D) = D\), or \(1/\lambda_{\rm 1st}(D)=1/D\). This is a discontinuous transition between $m=1$ and $m=0$. What drives the transition to continuity below \(D_{\mathrm{tri}}\) is the flip term, which promotes the emergence of the third state \(s_x=-1\).

\subsection{The two universality classes along the transition line}

\begin{figure}[ht]
 \centering
        \includegraphics[width=0.65\textwidth]{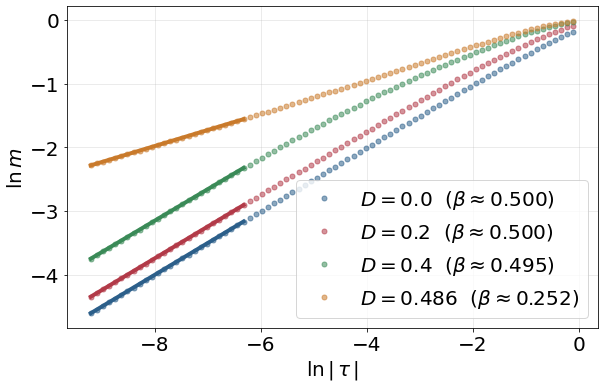}
 \caption{Log-log plot of the magnetisation vs $\tau$ and determination of the critical exponents $\beta_{\rm MFT}$ along the critical line and $\beta_{\rm tri}$ at the tricritical point. The fit of the data is performed on the left part of the plot (full lines). }
\label{fig3}
\end{figure}

It is instructive to check that the universality class remains that of Ising criticality all along the transition line, except at the tricritical point. For that purpose, we plot in Fig.~\ref{fig3} the behaviour of $\ln m$ vs $\ln |\tau|$, where straight lines with slopes $\beta$ are expected since we assume the power laws 
$m\sim|\tau|^\beta$ in the critical region $\tau\to 0^-$.

 At three values $D=0$, 0.2 and 0.4,  the respective slopes (0.500, 0.500, 0.495) agree with $\beta_{\rm MFT}=\frac 12$.  At the tricritical point $D=D_{\rm tri}$, the slope (0.252) agrees with  $\beta_{\rm tri}=\frac 14$.

\newpage

\begin{figure}[ht]
 \centering
        \includegraphics[width=0.65\textwidth]{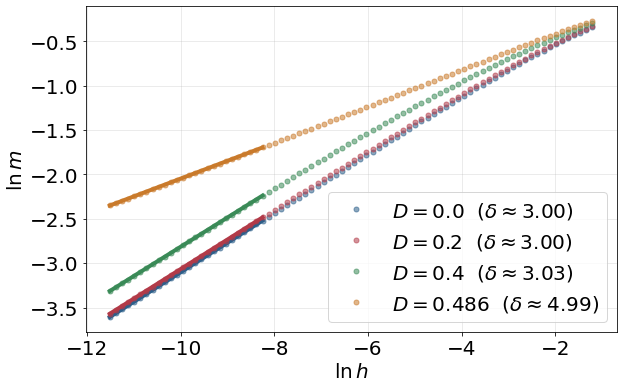}
 \caption{Log-log plot of the magnetisation vs $h$ and determination of the critical exponents $\beta_{\rm MFT}$ along the critical line and $\beta_{\rm tri}$ at the tricritical point. The fit of the data is performed on the left part of the plot (full lines).}
\label{fig4}
\end{figure}

In Fig.~\ref{fig4} we proceed the same way for the magnetic sector, testing now $m\sim |h|^{1/\delta}$ i.e. we plot the behaviour of $\ln m$ vs $\ln h$ for positive $h$ at $\tau =0$ -- that is at the respective critical values $\lambda_c(D)$ -- for three values $D=0$, 0.2 and 0.4, where the resulting slopes (0.333, 0.333, 0.330) agree with $\delta^{-1}_{\rm MFT}=\frac 13$ ($\delta_{\rm MFT}=3$). At the tricritical point $D=D_{\rm tri}$, the slope is $0.200$, in agreement with $\delta_{\rm tri}^{-1}=\frac{1}{5}$ ($\delta_{\rm tri}=5$).

\newpage

\subsection{Lee-Yang zeros distribution in the complex plane}

The study of phase transitions can be pursued in a somewhat more formal way by considering the zeros of the partition function in the complex plane. This method of investigation was introduced in the 1950s by Lee and Yang in two seminal papers~\cite{YangLee1952_I,LeeYang1952_II} and is presented e.g. in Goldenfeld~\cite{Goldenfeld1972LecturesOP}. 
This technique was used for the $\phi^4$ model by Kenna and Lang~\cite{Kenna1991} and recently in the mean-field regime~\cite{wada2026leeyangzerosedgesingularity}.
It is based on the fact that, for a finite system, the partition function is a finite sum of exponentials, which can be written as a polynomial in terms of the fugacity $e^{-\beta h}$. Since any polynomial of degree $N$ has exactly $N$ complex roots, extending the magnetic field to complex values causes the zeros of the partition function to distribute themselves in the complex plane and to converge, when the temperature is fixed to $T=T_c$ --  here the coupling $\lambda=\lambda_c$ -- towards the real critical point $h=0$ in the thermodynamic limit.

These zeros govern the singular behaviour of the free energy, and their study -- including their scaling, density, and distribution in the complex plane -- provides valuable information about the universality class of the transition. Here, we will highlight the significant differences between the second-order critical line and the tricritical point.

The object we need to study is not the variational partition function $Z(h)=e^{-\epsilon_1(h)}$. In the presence of a complex magnetic field, the eigenvalues of the Blume–Capel
\be
N^{-1}(\op H_{\rm MF})_{s_x}=\begin{pmatrix}
D-2\lambda m -h & -2^{-1/2} & 0 \\
-2^{-1/2} & 0 & -2^{-1/2} \\
0 & -2^{-1/2} & D+2\lambda m +h
\end{pmatrix}
+ \lambda m^2
\begin{pmatrix}
1 & 0 & 0\\
0 & 1 & 0\\
0 & 0 & 1
\end{pmatrix}
\ee
  become complex, but $e^{-\epsilon_1}$ itself can never vanish, since its modulus is always strictly positive. 
   
  The zeros of the partition function must therefore arise from a different mechanism. In the complex-field plane, the magnetisation also becomes complex and can take several saddle-point values. It is the destructive interference between the contributions associated with these different complex saddle points that produces the zeros of the $N$ mean-field-coupled spins partition function
\be
Z_N(\lambda,h)=\int dm\, e^{-N\epsilon_1(m,\lambda,h)}.
\ee

There is consequently no need to employ Cardano's method to determine the eigenvalues explicitly. We can instead work directly at the level of the Landau expansion and recalculate the coefficients $b_1(0)$, $b_1'(0)$, and so on, in a complex magnetic field. Right at the critical coupling, the coefficient of the quadratic term vanishes, and the expansion \eqref{Eq_LandauExp} in the presence of a magnetic field takes the form
\be f(m,h)-f_0=\tfrac 1n u_n m^n - hm,\ee
with $n=4$ along the critical line and $n=6$ at the tricritical point
with the two stable phases given by the saddle-point solutions of $h=u_nm^{n-1}$, i.e. $m_\pm=\pm m_0$.
These two competing phases have free energies $f_\pm=f_0+ \tfrac 1n u_n m_0^n\mp hm_0$. In a complex magnetic field $h=|h|e^{i\varphi}$, they contribute to the partition function with identical amplitudes,
\be
Z\simeq e^{-Nf_+}+e^{-Nf_-}.
\ee
A zero occurs when (see SM, sec. 3)
\be
\Re (f_+ -f_-)=\Re (-2m_0h)=0,
\ee
hence $\Re(h)=0$, or
\be
\varphi_{\rm LY}=\frac\pi 2,
\ee
whether $n$ is equal to $4$ or $6$. 
This result, once translated into the fugacity $e^{-h}-$complex plane, corresponds to the Lee--Yang circle theorem~\cite{Itzykson:1983zz,Itzykson_Drouffe_1989,Krasnytska_2016,kozitsky2026leeyangpropertyblumecapelmodel}, which has been extended to general quantum Ising models with complex many‐spin interactions in Ref.~\cite{10.1063/1.1665583}.
 
\begin{figure}[ht]
 \centering
        \includegraphics[width=0.45\textwidth]{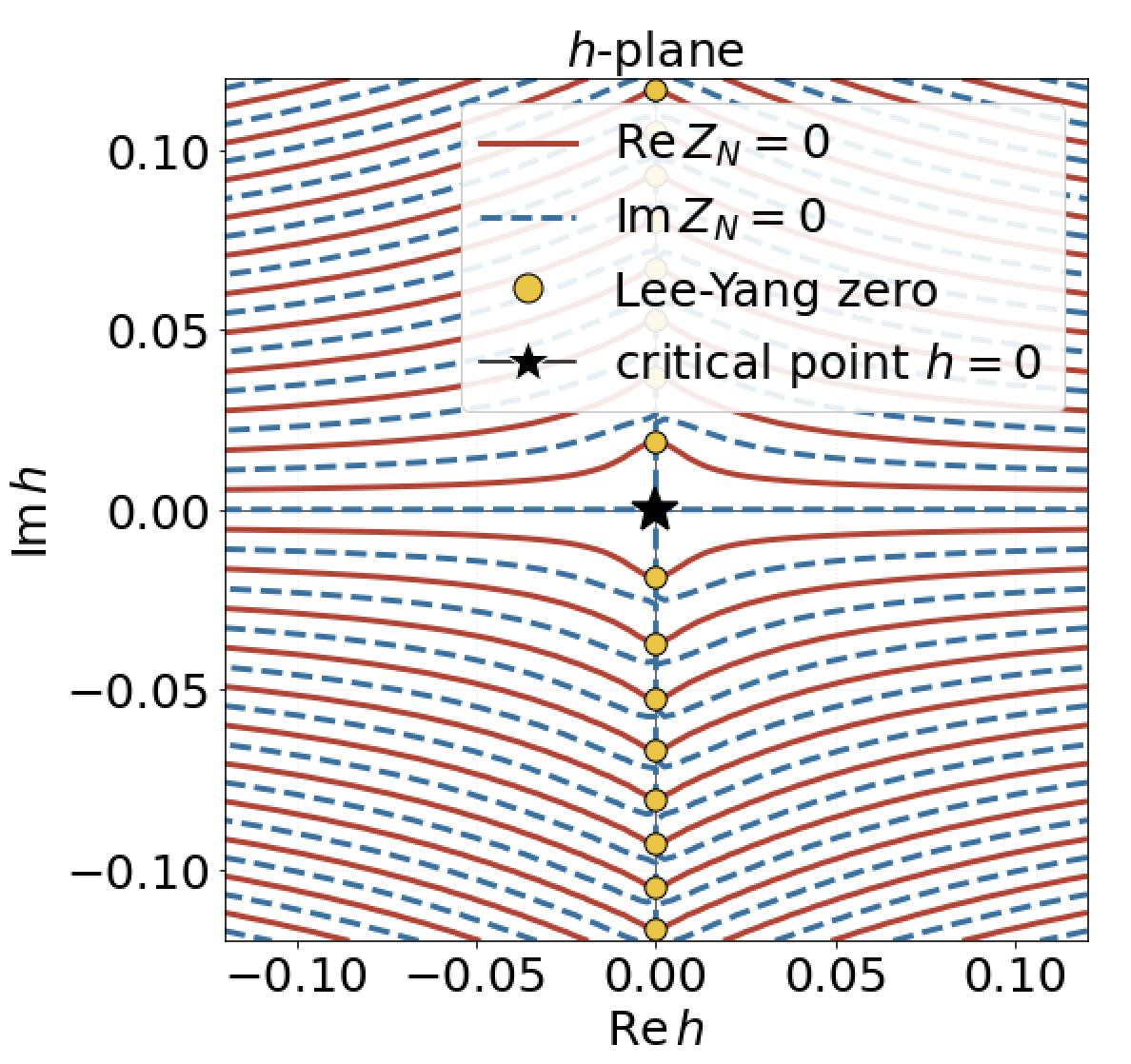}        \includegraphics[width=0.45\textwidth]{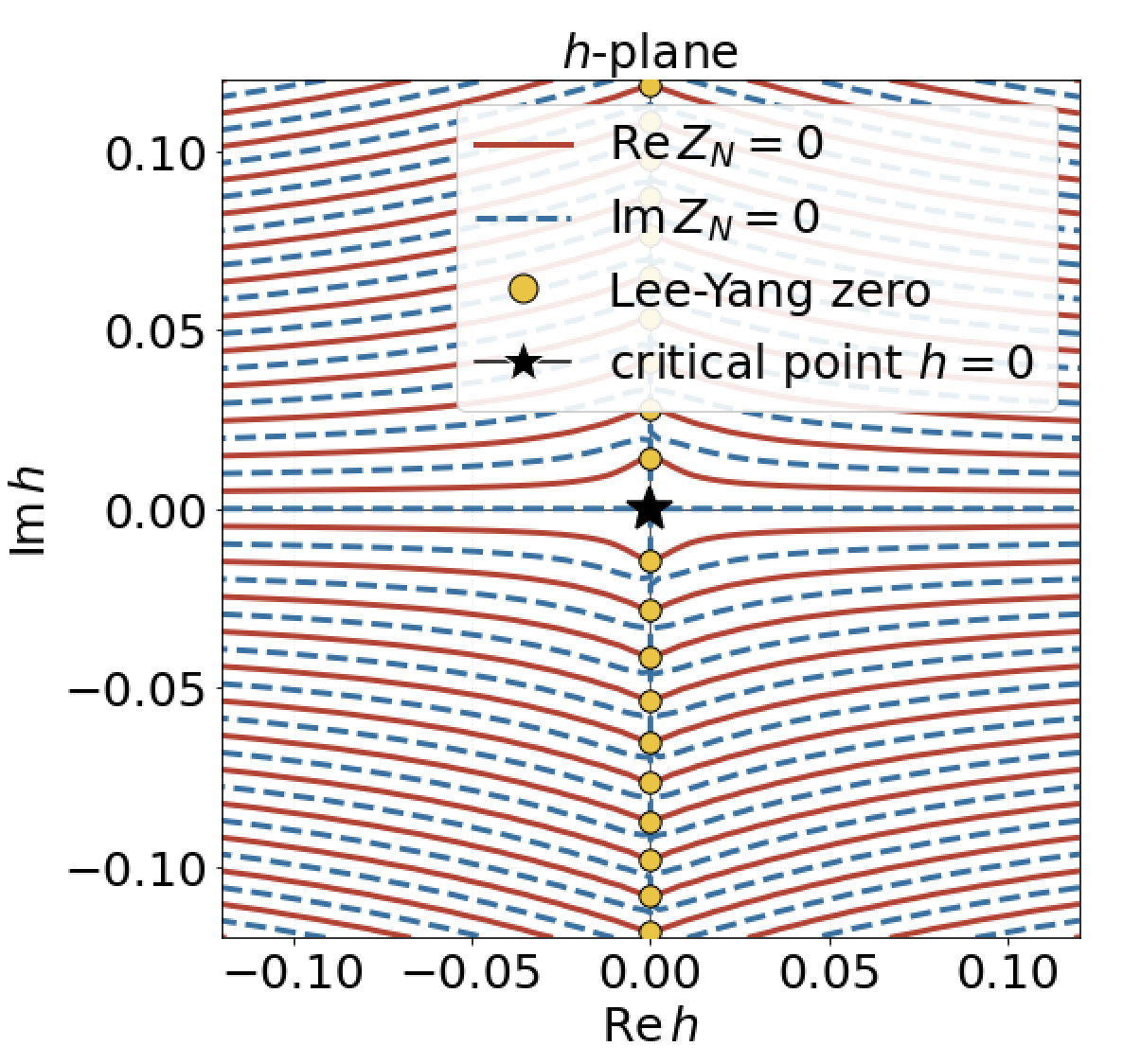}

 \caption{Distribution of the zeros of the partition function in the magnetic field complex plane with $N=500$ (left: second-order transition $D=0.2$, right: tricritical point $D=D_{\rm tri}$). The solid lines (red online) are the loci of zeros of $\Re(Z)$ and the dashed lines (blue online) those of $\Im(Z)$. The circles are the Lee--Yang zeros, and the central star corresponds to the phase transition.}
\label{fig5}
\end{figure}

 This is therefore a very robust result, which does not distinguish between the two regimes of continuous phase transitions, but deserves some attention.
Numerically,  for a system of size $N$, the mean-field partition function is obtained as an integral over the real variable $m$
\be Z_N \propto \int_{-\infty}^{+\infty} d \exp\big[-N f(m,\lambda,h)\big],\ee
with only $h$ (Lee-Yang zeros) or $\lambda$ (Fisher zeros, see next subsection) being analytically continued into the complex plane.
$Z_N(\lambda_c,h)$ (Lee--Yang zeros) is evaluated by direct numerical integration of the above integral over a fine grid of real $m$ values (typically $n_m\sim 4\times10^3$ points over $[-3,3]$), rather than by a saddle-point approximation. The zeros of $Z_N(h)$ are then located as the geometric intersections of the level curves $\Re (Z_N)=0$ and $\Im (Z_N)=0$.
We illustrate the results in the second-order regime and at the tricritical point in Fig.~\ref{fig5}, where the impact angle $\pi/2$ is found in the magnetic field complex plane (more is presented in the SM, sec. 3, in particular the first-order regime).

From this figure, there does not seem to be a significative difference between the two universality classes.

\subsection{Fisher zeros distribution in the complex plane}
Fisher has extended the partition function zeros to the complex temperature plane~\cite{FisherZeros}, here, the complex plane of the $\lambda$ parameter.
The numerical strategy is the same. Competing saddles come from the complex roots of
\be 
r(\lambda)+u_4(\lambda) m^2+ u_6(\lambda)m^4=0,
\ee
with $r(\lambda) =2\lambda\Bigl(1-\frac{\lambda}{\lambda_c(D)}\Bigr)\simeq 2\lambda_c^2\tau$
 (remember that the quantum control parameter is  $\tau=\lambda^{-1}-\lambda_c^{-1}$). Treating the two regimes at once like we did before, we write,  
 \be
 f(m,\tau)-f_0=\tfrac 12\lambda_c\tau^2+\tfrac 1n u_nm^n.
 \ee
 The saddle-point solutions are respectively $m=0$ for $\tau>0$ and $m=[\frac{2\lambda_c^2}{u_n}(-\tau)]^{\frac{1}{n-2}}$ for $\tau<0$, leading to a free energy difference
 \be 
\Delta F=\Bigl(\frac 1n-\frac 12 \Bigr)\Bigl(\frac{u_n}{2\lambda_c^2}\Bigr)^{\frac{2}{2-n}}
(-\tau)^{\frac{n}{n-2}}.
 \ee
 Setting a complex $\tau=|\tau| e^{i\varphi}$, the condition for a Fisher zero is therefore 
 \be
 \cos\Bigl(\frac{n\varphi}{n-2}\Bigr)=0
 \ee
 leading to 
 \be
 \varphi_{\rm F}=\frac{n-2}{2n}\pi
 \ee
 or $ \varphi_{\rm F}=\pi/4$ for a second-order mean-field transition and $ \varphi_{\rm F}=\pi/3$ for a tricritical point in the mean-field theory.

\begin{figure}[ht]
 \centering
        \includegraphics[width=0.45\textwidth]{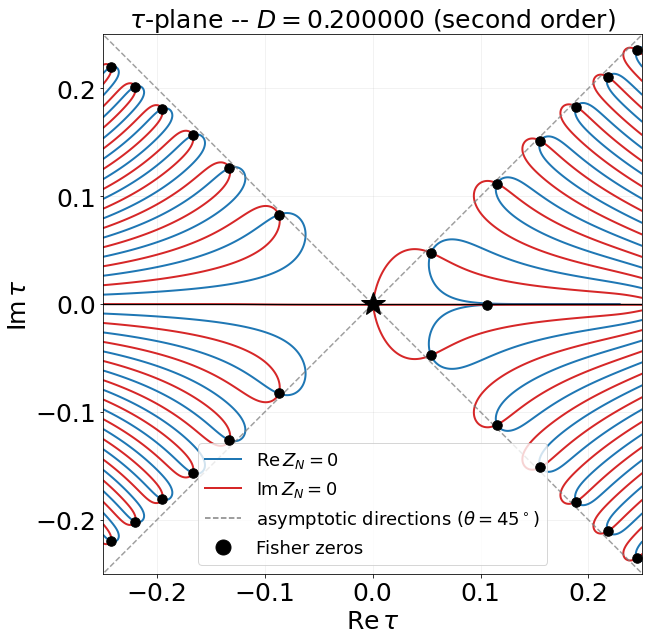}   \includegraphics[width=0.45\textwidth]{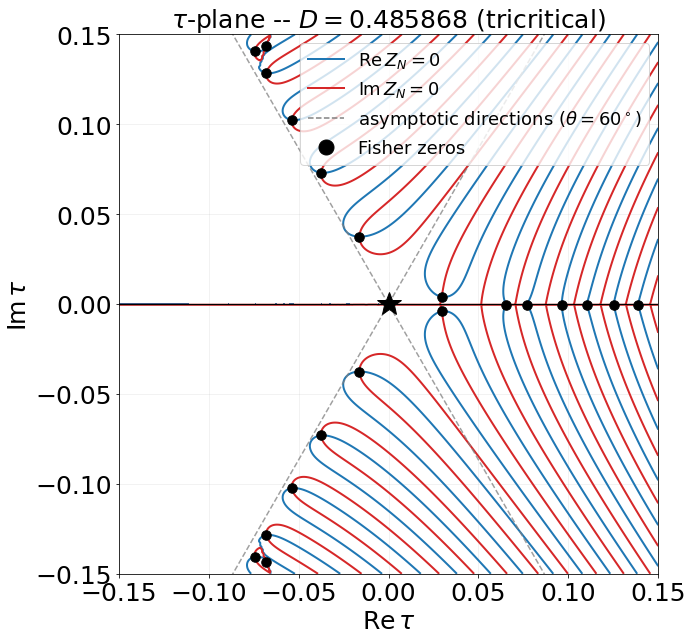}

 \caption{Distribution of the zeros of the partition function in the coupling complex plane obtained here with $N=1000$. The solid lines (red online) are the loci of zeros of $\Re(Z)$ and the dashed lines (blue online) those of $\Im(Z)$. Left panel: second-order transition $D=0.2$, the Fisher impact angle measured from the negative temperature real part is equal to  $\pi/4$. Right panel: tricritical point $D=D_{\rm tri}$, the Fisher impact angle is equal to $\pi/3$). }
\label{fig6}
\end{figure}

To proceed numerically, at $h=0$, we use the symmetry $m\to-m$ of $f$ which reduces the set of real stationary points to three: the central saddle point $m=0$ and a symmetric pair $\pm m^*(\lambda)$. $Z_N(\tau,h=0)$ is then approximated by the two-term saddle-point sum, which is physically motivated by the coexistence of two phases:
$$Z_N(\tau,h=0)\approx A_0,e^{-Nf_0} + 2A_1,e^{-Nf_1(t)},$$
with $A_s=\sqrt{2\pi/(N f_s'')}$. The physical branch is propagated continuously across the $\tau$ grid by means of a breadth-first search starting from a ring of seed points surrounding the excluded region around $\tau=0$ -- a treatment that is necessary because $f_0$ and $f_1$ become degenerate as the critical point is approached. Candidate zeros, located as above by intersecting the level curves, are then refined by a two-dimensional Newton iteration on $\Re(Z_N)$ and $\Im(Z_N)$.

This is confirmed in Fig.~\ref{fig6}.
The patterns are very different in the two cases, and this is a spectacular signature of the difference in universality classes.

Note that in the general case~\cite{Itzykson:1983zz,krasnytskaPhD,Krasnytska_2016}, the formula for $\varphi_{\rm F}$ involves the exponent $\alpha$ and the ratio of critical amplitudes for the specific heat.

\section{Conclusion}
 In this paper, we have demonstrated how the mean-field approach can be introduced to study critical behaviour in a quantum chain. Although mean-field theory is one of the most widely used theoretical tools and is routinely employed as a first approach to understanding the properties of a physical system, its application to quantum phase transitions is much less common. Yet, we believe that it provides a particularly instructive and pedagogical framework, making the analogies with thermally driven phase transitions especially transparent.

The Blume--Capel model constitutes an ideal framework for such a study. As in its classical counterpart, it displays a remarkably rich phase diagram, featuring a continuous phase-transition line, a tricritical point, and a first-order transition line. Moreover, it is amenable to both exact analytical and perturbative treatments, while remaining sufficiently simple to be accessible to undergraduate students. We therefore believe that it provides an excellent pedagogical example for introducing the fundamental concepts of quantum critical phenomena.

In conclusion, it is worth noting that mean-field theory and Landau theory -- as specified in the introduction -- are merely approximate methods for describing phase transitions, valid in principle above the upper critical dimension. However, it is known that even in this regime, they do not provide a complete understanding due to the presence of dangerous irrelevant variables~\cite{FisherStellenbosch}. This aspect is now well understood and has been extensively studied for many types of phase transitions~\cite{SciPostPhysLectNotes.60,PhysRevLett.89.025703,kenna2017universal,Holovatch2015Volume4,Flores_Sola_2015}, including -- more recently -- quantum phase transitions ~\cite{SciPostPhys.13.4.088}; interested readers may refer to the excellent review~\cite{e26050401}.

\ack{LM and BB wish to pay special tribute to their friend, the late Ralph Kenna, who introduced them to the fascinating world of the zeros of the partition function.
The authors ``thank'' Claude (Anthropic, Sonnet 4.5 model) for assistance in developing the Python scripts used in this work, in particular for the numerical solution of the self-consistency equations with {\tt brentq} and the production of the figures, those of the zeros of the partition function in particular. All numerical results were independently verified by the authors. }

 
\newpage


\input 25_MFT_QuantumChains.bbl

\end{document}

%% file: 25_MFT_QuantumChains.bbl
\providecommand{\newblock}{}